\documentclass{aa} 
\usepackage{graphicx}
\usepackage{mathabx}
\usepackage{hyperref}
\hypersetup{colorlinks,allcolors=black}
\usepackage{txfonts}

\usepackage{amstext}

\begin{document}

   \title{The best place and time to live in the Milky Way}

\titlerunning{The best place to live in the Milky Way}
\authorrunning{R.~Spinelli et al.}
   \author{R.~Spinelli \thanks{email:r.spinelli@studenti.uninsubria.it}\inst{1,2}, G.~Ghirlanda\inst{2},  F.~Haardt\inst{1,2,3}
   G.~Ghisellini\inst{2}, G. Scuderi\inst{4}}
   \institute {Dipartimento di Scienza e Alta Tecnologia, Universit\`a dell'Insubria, Via Valleggio 11, 22100 Como, Italy
    \and  
 INAF -- Osservatorio Astronomico di Brera, Via E. Bianchi 46, 23807 Merate (LC), Italy
 \and
 INFN -- Sezione Milano--Bicocca, Piazza della Scienza~3, 20126 Milano, Italy
 \and 
  Dipartimento di Fisica G. Occhialini, Universit{\`a} Milano--Bicocca, Piazza della Scienza~3, 20126 Milano, Italy
 }

   \date{Received September 18, 2020; accepted }

  \abstract
   {Counted among the  most powerful cosmic events, supernovae (SNe) and $\gamma-$ray bursts (GRBs) can be highly disruptive for life: Their radiation can be harmful for biota or induce extinction by removing most of the protective atmospheric ozone layer from terrestrial planets (TPs). Nearby high-energy transient astrophysical events have been proposed as  possible triggers of mass extinctions on
   Earth.}
   {We assess the habitability of the Milky Way (MW) throughout its cosmic history against potentially disruptive astrophysical transients with the aim of identifying the safest places and epochs within our Galaxy. We also test the hypothesis that one long GRB played a leading role in the late Ordovician mass-extinction event ($\sim 445$ Myr ago).}
   {We characterised the habitability of the MW throughout its cosmic history as a function of galactocentric distance of TPs. 
   We estimated the dangerous effects of transient astrophysical events (long and short GRBs and SNe) with a model that connects their rate to the specific star formation and metallicity evolution within the Galaxy throughout its cosmic history.  
   Our model also accounts for the probability that TPs
form   around FGK and M stars.}
   {
   Until $\text{about six}$ billion years ago, the outskirts of the Galaxy were the safest places to live, despite the relatively low density of TPs. In the last $\text{about four}$ billion years, regions between 2 and 8 kpc from the center, which had a higher density of TPs, became the best places for a relatively safer biotic life growth. We confirm the hypothesis that one long GRB played a leading role in the late Ordovician mass-extinction event. In the last 500 Myr, the safest neighborhood in the Galaxy was a region at a distance of 2 to 8 kpc from the Galactic center, whereas the MW outskirts were sterilized by two to five long GRBs. 
   }
   {}

   \keywords{Gamma-ray-burst: general, Galaxy: evolution, Astrobiology
               }

   \maketitle
%

\section{Introduction}

One of the primary goals of exoplanetary research is to find habitable worlds. 
In order to assess the very notion of "habitable", we  
must rely on our understanding of the reasons for the presence and absence of lifeforms in the Solar System. 
Several factors determined the appearance and development of life on planet Earth. 
In addition to particular intrinsic properties of the planet 
(e.g., geology and magnetic field) and solar characteristics 
(e.g., spectrum and irradiation), it is understood that a key 
requirement for the development of life on Earth is the presence 
of liquid water on the planetary surface. 
Potentially habitable exoplanets are identified based on 
their location within the circumstellar habitable zone \citep[CHZ; e.g.,][]{1993Icar..101..108K,2013ApJ...765..131K}. 

In addition to local factors, planetary habitability could also be  affected by the galactic environment, for example, by astrophysical events outside the Solar System that can irradiate the planet.
As many studies suggest \citep[e.g.,][]{1974Sci...184.1079R, 1995ApJ...444L..53T, 1998PhRvL..80.5813D, 2003ApJ...585.1169G, 2011AsBio..11..343M, 2012MNRAS.423.1234S}, high-energy transients such as supernovae (SNe) and 
gamma-ray bursts (GRBs) could be life-threatening and a potential cause of
mass extinctions. 
A GRB, with a typical isotropic equivalent energy of $10^{52}$ erg located 
within $\sim$1 kpc from the Earth, would irradiate its atmosphere with a 
$\rm \gamma$-ray (i.e., keV--MeV) fluence $\ge$100 kJ m$^{-2}$ ($10^8$ erg cm$^{-2}$). 
This level of irradiation 
can produce  stratospheric nitrogen compounds, which quickly destroy 90\% of the ozone layer on average \citep{2005ApJ...622L.153T}. 
As a first consequence, the higher solar UVB radiation that would reach the surface of Earth would 
be harmful to life. 
Intense UVB radiation could also be lethal to surface marine life such as
phytoplankton, which is crucial for the food chain and oxygen production. 
Moreover, the opacity of the NO$_2$ produced in the stratosphere would reduce 
the visible sunlight that reaches the surface, causing a global cooling. As argued 
by \citet{Herrmann2002} and \citet{2003Geo....31..485H}, the late Ordovician mass-extinction event 
($\sim$445 Myr ago), which is one of the five great mass extinctions on Earth, 
has some climatic signatures that can be interpreted by invoking an extra-terrestrial 
cause such as a nearby GRB \citep{2005GeoRL..3214808M}. 

The lethality of transient astrophysical events depends on their energy released as high-energy radiation and their occurrence rate in the Galaxy: more powerful events can be lethal for a planet over larger distances, while a high event rate can also reduce the ability of the planet to recover 
from the environmental effects induced by the radiation of the event. \citet{2014PhRvL.113w1102P} and \citet{2015ApJ...810...41L} 
consistently found that long-duration GRBs (with an observed 
duration $>$2 s; LGRBs hereafter) are the most dangerous 
astrophysical events for the Earth, even more so than short-duration 
GRBs (lasting $<2$ s; SGRBs hereafter) and SNe. 
This is mainly due to the high energy $10^{51-54}$ ergs 
(isotropic equivalent) released by LGRBs, which
compared to SNe compensates for their lower intrinsic rate 
($\sim 5 \times 10^{-6}$ yr$^{-1}$ per galaxy according to \citealt{2010MNRAS.406.1944W}). In particular, the probability is non-neglibile (50\% according to \citealt{2014PhRvL.113w1102P}) 
that in the last 500 Myr the Earth could have been illuminated 
by one long lethal GRB (precisely $\sim$0.93 according to \citealt{2015ApJ...810...41L}).

The rate of  astrophysical events is linked to the properties (and their
variation with cosmic time) of the environment in which they occur. 
The rate of LGRBs, which is connected to the end life of 
massive stars \citep[e.g.,][]{1993ApJ...405..273W}, is expected to be 
proportional to the efficiency of converting gas into stars 
(i.e., the star formation rate, SFR). 
However, their progenitors are thought to have a low metallicity to conserve 
the angular momentum required to  efficiently launch the jet \citep{2006ApJ...637..914W, 2006A&A...460..199Y}. 
This argument roughly agrees with the observed preference 
of LGRB to occur in relatively low-metallicity 
host galaxies \citep[e.g.,][]{2016A&A...590A.129J,2019A&A...623A..26P}. When the evolution of the rate of LGRBs 
within the Galaxy is computed, the possible evolution 
of the SFR and the gas metallicity with cosmic time and within the Galaxy therefore need to be accounted for. 

The ever-increasing number of extrasolar planets ($\sim$4330 at present)
motivates the extension of these studies to the whole Galaxy. The outer regions of the MW appear to be the most 
favorable for preserving life \citep{2014PhRvL.113w1102P,2015ApJ...810...41L,2016MNRAS.459.3512V} 
because the SFR is reduced. 
However, \citet{2014PhRvL.113w1102P} and \citet{2015ApJ...810...41L} 
scaled the cosmological rate of LGRBs in proportion to the stellar 
mass of the MW disk, assuming constant metallicity and a specific 
star formation rate (sSFR). In this work we account for the radial 
distribution and the inside-out evolution of metallicity and 
SFR within the Galaxy.

\citet{2011AsBio..11..855G}, \citet{2014MNRAS.440.2588S}, and \citet{2016MNRAS.459.3512V} defined a Galactic habitable zone (GHZ) 
by considering only SNe as possible deleterious events and focused 
on TPs orbiting FGK stars. 
It is worth extending these studies by considering GRBs and M stars, 
which are the most powerful astrophysical events and the most abundant 
stellar population in the MW, respectively.

We examine the astrophysical constraints for life in the 
MW and for the first time consider all the most energetic transient 
events (SGRBs, LGRBs, and SNe). We link their rates to the 
sSFR within the Galaxy and to its variation with 
cosmic time through a semianalytical model describing the evolution of the MW.
For LGRBs we consider the variation of the metallicity of the MW.
Finally, with a similar method as was used by \citet{2017A&A...605A..38S}, 
we account for the probability that TPs 
(also dependent upon metallicity) form around FGK and M stars 
\citep{2016ApJ...833..214Z}. Throughout the cosmic history of the MW we identify the safest (from an astrophysical perspective)
locations. These are sites suitable for the presence of 
planets with long-lasting biospheres. 

Our work is organized as follows. In \S \ref{sec:2} we present
the methods for estimating the number of lethal events as a function of 
the position within the Galaxy and of the cosmic time. 
In \S \ref{sec:3} we  present model assumptions and the model we adopted to compute the 
evolution of the star formation, metallicity, and planetary density
within the MW. 
In \S \ref{sec:5} and \S \ref{sec:6} we present and discuss our results.
In \S \ref{sec:conclusion} we summarize our results. 
We adopt a $\Lambda$CDM cosmological model with  $\Omega_{\rm M}= 0.3 $, 
$\Omega_\Lambda=0.7,$ and $H_0=70$ km s$^{-1}$ Mpc$^{-1}$. 

\section{Methods}
\label{sec:2}


\subsection{Cosmic rate}
For a generic population of astrophysical sources described by a function 
$\xi$ in the luminosity-redshift ($L-z$) space, the cosmic rate (number of events per unit comoving volume and time) at any cosmic epoch is
\begin{equation}
    \frac{dN}{dV dz} =\int_{L} \xi(L,z) dL
.\end{equation}
%

\subsection{Scaling down to the Milky Way}

In order to estimate the rate of a population of  astrophysical sources within 
the MW, we rescaled its known cosmic rate at any cosmic epoch within the 
cosmological volume occupied by the Galaxy, 
\begin{equation}
     \frac{d \mathcal{N}_{\rm MW}(z)}{dz} = \int \xi (L,z) V_{\rm MW} (z) \mathcal{P} (z) \ dL 
,\end{equation}
where V$_{\rm MW}(z)={M_{\star}(z)}/{\rho_{\star}(z)}$ is the cosmological 
volume occupied by the MW at a given redshift, 
$\rho_{\star}(z)$ is the average stellar density as a function of 
redshift $\rho_{\star}(z) = 10^{17.46-0.39z}$ M$_{\odot}$ Gpc$^{-3}$ \citep{2015ApJ...810...41L, 2015MNRAS.447....2M}, and $M_{\star}(z)$ 
is the stellar mass of the evolving Milky Way (\S 3.2). 
$\mathcal{P} (z)$ is the probability that astrophysical sources (i.e., LGRB, SGRB, and SN) occur within the MW at a given cosmic epoch (the cosmic time is here represented as the redshift $z$). 
This probability depends on the cosmic evolution of the MW properties 
(e.g., sSFR and metallicity in the case 
of LGRBs), which can inhibit or favor the occurrence of the lethal transient sources
under consideration.

\subsection{Rate of lethal events within the MW}

The fluence produced by astrophysical transients in  a planetary atmosphere 
is the primary ingredient leading to possible lethal effects. \citet{2005ApJ...622L.153T,2005ApJ...634..509T} estimated in a 
2D atmospheric model that a $\gamma$--ray fluence of 10 kJ m$^{-2}$ 
can on average induce a 68\% depletion of the ozone layer at an 
altitude of 32 km on a timescale of a month. Higher fluences, 
for instance, 100 kJ m$^{-2}$ and 1000 kJ m$^{-2}$ , would produce depletions 
up to 91\% and 98\%, respectively. 
We consider a depletion of 91\% of the ozone layer sufficient to produce 
mass extinctions (see also \citealt{2005ApJ...622L.153T,2005ApJ...634..509T,2015ApJ...810...41L}). 
We therefore define as astrophysical lethal events that are capable of
illuminating a planetary atmosphere with a fluence (i.e., energy flux 
integrated over the event duration) F$\ge$100 kJ/m$^2$ 
(i.e., 10$^8$ erg cm$^2$, F$_c$). 

Given a population of astrophysical events, the lethal effect on a planet can be quantified by computing the rate of lethal events. 
At any cosmic time, the rate per unit time of lethal events (i.e., with a fluence $\ge F_c$) for a planet at distance R from the Galactic center is 
\begin{equation} \label{eq:3}
     \frac{d \mathcal{N}_{\rm MW}(R, z)}{d z}  = \int \xi (L,z) V_{\rm MW} (z) \mathcal{P} (d, z | R) \ dL 
,\end{equation}

where $\mathcal{P}(d,z \mid R)$ (see \S 2.4) is the portion in mass of the Galaxy contained within the region where an event with energy E is lethal for a planet at R (i.e., the portion of the MW within a distance $d$ from $R$) and describes the probability that lethal events occur (i.e., with an energy and distance producing a fluence  $\ge$ $F_c$) within the MW given 
its local properties. The integral is performed over the entire luminosity distribution. 

\subsection{Portion of the Galaxy}

We define the hazard distance $d(E,F_c)$ of an astrophysical event with energy $E$ as the (lethal) distance within which its fluence is higher than $F_c$,
\begin{equation}
d(E,F_c) = \sqrt{\frac{E}{4 \pi F_c}} 
\label{eq:4}
.\end{equation}
\smallskip
Assuming that GRBs and SNe follow the stellar distribution within the Milky Way, $\mathcal{P}(d,z \mid R)$ at a given time t can be calculated by integrating the MW stellar surface density $\Sigma_{\star}(R,z)$ (see \S \ref{sec:4.2}) within a distance $d$ from the position of the planet ($R$), 
\begin{equation}
    \mathcal{P}(d,z \mid R) = \frac{1}{M_{\star} (z)} \int_{S} \Sigma_{\star} (R,z) \ da
\label{eq:5}
.\end{equation}
We adopt a polar coordinate system (see Fig. 1), centered at the position 
at distance $R$ from the Galaxy center, to calculate $ \mathcal{P}(d,z \mid R)$,
\begin{gather*}
     \mathcal{P}(d,z \mid R) = \frac{d(L,F_c)^2}{M_{\star} (z)} \int_{0}^{2\pi} d\phi \int_{0}^{1} \mu \ \Sigma_{\star} (q,z) \ d\mu \\
    \\
    q = \sqrt{R^2 + r^{'2} + 2d \ R\mu \ cos(\phi)} \\
    r'\equiv \mu d  \ \ \ \ \text{with} \ \ \ 0 \le \mu \le 1.
\end{gather*}
%
%
 \begin{figure}
   \centering
   \includegraphics[scale=0.16]{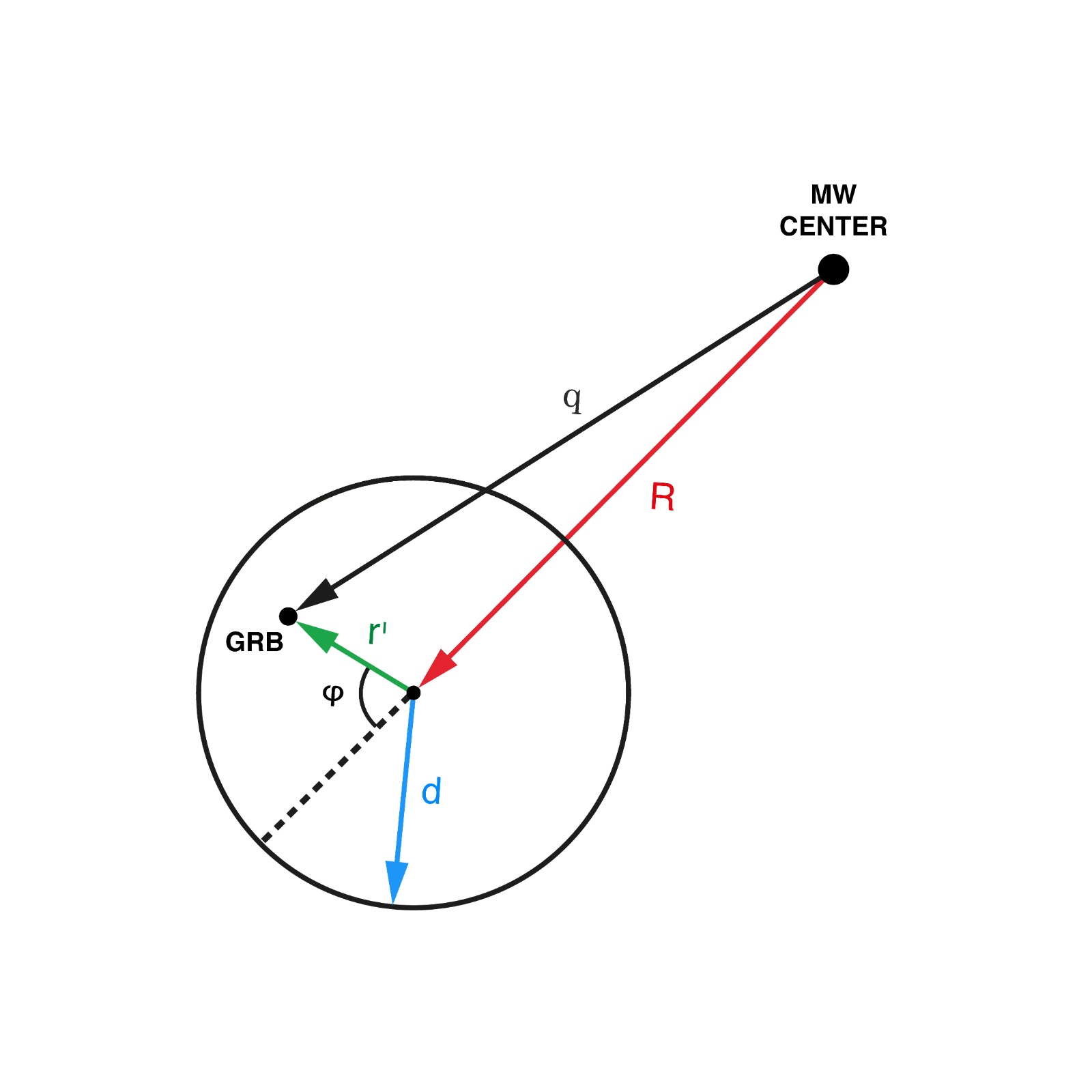}
   \caption{Portion of the Galaxy where an event with energy E is lethal for a planet at R (solid black circle). 
   The hazard distance $d$ identifies the circular region within which a 
   GRB of energy $E\ge 4\pi d^2 F_c $ can produce a lethal fluence. 
   $q$ represents the galactocentric distance of the GRB. 
   In this sketch the center of the polar coordinate system 
   (identified by the arrow $R$) is the location of the planet. } 
    \label{portion}%
    \end{figure}
%

\section{Model assumptions} \label{sec:3} 

For GRBs 
and SNe it is customary to factorize $\xi(L,z)=\phi(L)\psi(z)$. Here 
$\phi(L)$ describes 
the luminosity distribution of the event rate at $z=0$, while the dimensionless function $\psi(z)$ is its redshift evolution.

\subsection{Luminosity function}

We assume a broken power law $\phi(L)$ for LGRBs  
\citep{2010MNRAS.406.1944W, 2012ApJ...749...68S, 2016A&A...587A..40P} 
and for SGRBs 
\citep{2005A&A...435..421G, 2010MNRAS.406.1944W,  2014MNRAS.442.2342D,  2016A&A...594A..84G}, defined between $L_{\rm min}$ and $L_{\rm max}$,
\begin{equation} \label{eq:6}
\phi(L)=n_0 
    \begin{cases}
      \left(\frac{L}{L_b}\right)^{-\alpha} & \text{if $L_{\rm min} < L < L_b$}\\
      \left(\frac{L}{L_b}\right)^{-\beta} & \text{if $L_b < L < L_{\rm max,}$}
    \end{cases}       
\end{equation}
where $n_0$ is the present-day (i.e., $z=0$) rate at the break luminosity $L_b$. GRBs are jetted sources, thus only the GRB jets pointing at the planet can be harmful. Here $n_0$ is the observed rate (not corrected for collimation) and L is the isotropic equivalent luminosity (see \S 3.4 for details).  n$_0$ is the ratio between $\rho$ (i.e., the cosmological rate at z=0, Table \ref{tab:1}) and the integral of $\phi(L)$ with unit normalization. We assume that the LGRB and SGRB distributions have a characteristic duration centered on $\tau$ of 20s and 2s, respectively \citep{1993ApJ...413L.101K}. The energy (necessary to calculate the fluence) is derived from the luminosity $L$ assuming that the burst with a duration $
\tau$ has a triangular shape. This is a fairly good approximation for  SGRBs but an oversemplification of the complexity of LGRB light curves. 
For long and short GRBs we adopt the parameter values reported in Table \ref{tab:1}.

\begin{table*} 
\centering
\begin{tabular}{cccccccc}
\hline
  \multicolumn{1}{c}{} &
  \multicolumn{1}{c}{$\rho$} &
 \multicolumn{1}{c}{$\alpha$}  &
  \multicolumn{1}{c}{$\beta$} &
  \multicolumn{1}{c}{$L_b$}&
  \multicolumn{1}{c}{$L_{min}$}&
  \multicolumn{1}{c}{$L_{max}$}&
  \multicolumn{1}{c}{$\tau$}
  \\
  &[Gpc$^{-3}$ yr$^{-1}$]
  &
  &
  &
  [ergs s$^{-1}$]&
  [ergs s$^{-1}$]&
  [ergs s$^{-1}$]&
  [s]
  \\
\hline
 LGRB & 1.3 $\pm$ 0.6 & 1.2 $\pm$ 0.9& 2.4 $\pm$ 0.77 & 10$^{\rm 52.5 \pm 0.2}$ & 10$^{\rm 49}$ & 10$^{\rm 54}$& 20\\
 SGRB &0.3 $\pm$ 0.06 & 0.53 $\pm$ 0.88& 3.4 $\pm$ 2.2& (2.8$\pm$2.1) $\times$ 10$^{\rm 52}$ & 5 $\times$ 10$^{\rm 49}$ & 10$^{\rm 53}$  & 2\\
\hline\end{tabular}
\medskip
\caption{Parameters of the broken power-law luminosity function of LGRBs and SGRBs \citep{2010MNRAS.406.1944W,2016A&A...594A..84G} and burst durations. $\rho$ is the cosmological rate at $z=0$. }
\label{tab:1}
\end{table*}

For SNe, we consider the cosmic rate at $z=0$  derived by  \citet{2012PASA...29..447M} and \citet{2011MNRAS.412.1441L}. The distribution of the energy output of SNe can be described as Gaussians \citep{1997MNRAS.290..360H,
1999A&A...351..459C, 2002AJ....123..745R, 
2004ApJ...602..571B, 2008A&A...479...49B, 2010AJ....139...39Y}
with parameter values reported in Table \ref{tab:2}. 
We further distinguish between the three different classes of SNe Ia, Ibc, and IIp.

\subsubsection{Redshift distribution}

\begin{figure}
   \centering
   \includegraphics[scale=0.45]{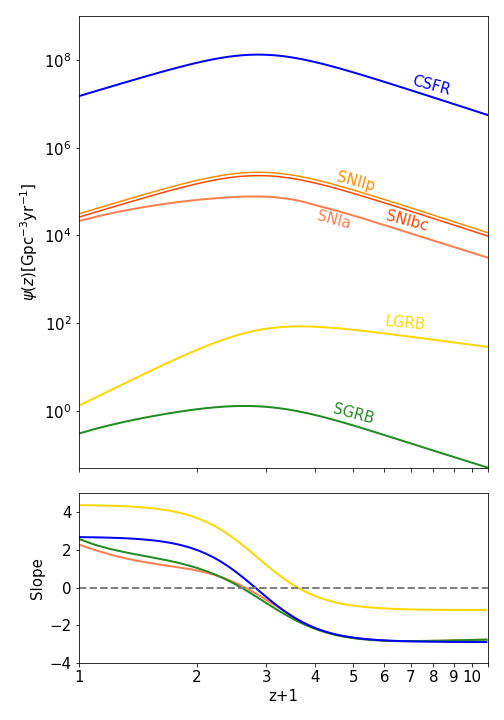}
   \caption{{\it Top panel}: Cosmic density rate of the three classes of high-energy transients considered: LGRB (yellow line), SGRB (green line), and SNe (orange lines). LGRB and SGRB rates are not corrected for the collimation angle, i.e., they represent the fraction of bursts whose jets are pointed toward the Earth. The CSFR (blue line) is in units of M$_{\odot}$ yr$^{-1}$ Gpc$^{-3}$. All the curves are normalized to their respective local rate (see Table \ref{tab:1} and Table \ref{tab:2} for GRBs and SNe, respectively). {\it Bottom panel}: Derivative of the formation rate curves shown in the top panel.  The color-coding is the same. 
   The derivative of SNIbc, IIp coincides with that of the CSFR (blue line in the bottom panel). 
   The horizontal line identifies for each curve the redshift $z$ corresponding to the maximum of the rate curve.} 
    \label{psi}%
\end{figure}

The association of LGRBs with envelope-stripped SNe 
\citep{1998Natur.395..670G, 2003ASAJ..113.1788S,2003Natur.423..847H,2005GCN..3087....1M,2006Natur.442.1011P,2006Natur.442.1008C,2011ApJ...735L..24S,2012A&A...547A..82M,2013ApJ...776...98X} and the properties of their hosts \citep[e.g.,][]{2006Natur.441..463F} 
probe their  origin from the core-collapse of rapidly 
rotating massive stars \citep{1993ApJ...405..273W, 1999ApJ...524..262M}. Because it is generated by the explosion of short-lived massive stars,  
the redshift distribution of LGRBs and SNIbc/IIp is expected to follow the
cosmic star formation history
(CSFR; e.g., \citealt{2014ARA&A..52..415M, 2006ApJ...651..142H}),
\begin{equation}
    \psi_{\star}(z) = 0.015 \frac{(1+z)^{2.7}}{1+[(1+z)/2.9]^{5.6}}\,\,\, \rm M_{\odot} yr^{-1} Mpc^{-3}
,\end{equation}
which is represented by the blue line in Fig. \ref{psi}. 
However, differently from SNIbc/IIp, the rate of LGRBs deviates from the CSFR  
\citep{2004ApJ...611.1033F,2006MNRAS.372.1034D,2006ApJ...642..636L,2007ApJ...657L..73G,2009ApJ...705L.104K,2011MNRAS.417.3025V,2012ApJ...749...68S}. 
This corresponds to a steeper (with respect to the CSFR) increase in the GRB rate with 
increasing redshift \citep{2016A&A...587A..40P} up to a peak corresponding to $z\sim3.5$ 
(i.e., higher than the CSFR peak at $z\sim2$).  
This could be interpreted as caused by the GRB bias (i.e., preference) 
for low-metallicity progenitors \citep{2006ApJ...637..914W}. 
Studies of the host metallicity have suggested that GRBs in most cases occur in galaxies 
whose metallicity $Z$ is lower than a threshold value $\sim 0.7 \, Z_{\odot}$ \citep{2018MmSAI..89..175V,2019A&A...623A..26P}. Population studies \citep{2017MNRAS.469.4921B, 2018NewA...65...73B} suggest that this metallicity threshold lies in the range 0.3--0.6  $Z_{\odot}$. 
Assuming a threshold value $Z_c = 0.4 \, Z_{\odot}$ \citep{1994A&AS..106..275B,2011MNRAS.417.3025V}, we therefore modeled the LGRB population under 
this hypothesis and express their cosmic rate (orange line in Fig. \ref{psi}) as
\begin{equation} \label{eq:8}
\psi_{\rm LGRB} (z) = \ \frac{\psi_{\star}(z) }{\psi_{\star}(0) }\frac{\Theta_{\rm Z<Z_c}(z)}{\Theta_{\rm Z<Z_c}(0)} \rm \,\,\,\, yr^{-1} Gpc^{-3}
,\end{equation}
where $\Theta_{Z<Z_c}(z) \rm$ is the  fraction  of  stars with a metallicity lower 
than $Z_c$. 
 We calculated $\Theta_{\rm Z < Z_c}(0)$ by assuming that the metallicity 
 of the local Universe has a  mean value [Fe/H]$_0$=-0.006 with a normal 
 dispersion $\sigma=0.22$ \citep{2008MNRAS.383.1439G,2014ARA&A..52..415M}. 
 As we show in \S 3.3, the final rate of LGRBs within the 
 MW does not depend on $\Theta_{Z<Z_c}(z),$ but only on its value at z=0.

SGRBs are thought to be produced by the mergers of compact objects, as 
recently proved by the multimessenger observations of the event GW/GRB170817 \citep{2017PhRvL.119p1101A,2017PhRvL.119n1101A}. 
It is expected that their redshift distribution does not directly follow 
the CSFR because of the delay  between their formation 
as a binary and their merger. The delay-time distribution is a power law with slope $-1$ 
between a few million and a few billion years  \citep{2005A&A...435..421G,2005AAS...20715803N,2006A&A...453..823G, 2015MNRAS.448.3026W,2011MNRAS.417.3025V}.
\citet{2016A&A...594A..84G} derived the SGRB formation rate from available observational constraints and found that it is indeed consistent with a delayed cosmic SFR history. 
We here adopt the parametric function obtained by their work 
(green line in Fig. \ref{psi}), 
\begin{equation}
    \psi_{\rm SGRB}(z)=\frac{1+2.8z}{1+(z/2.3)^{3.5}} \rm yr^{-1} Gpc^{-3}
.\end{equation}

For SNIa we assumed the redshift distribution derived by \citet{2012PASA...29..447M}. 
This function is derived by convolving the star formation history of 
\citet{2006ApJ...651..142H} with a power-law delay-time distribution (DDT $\sim t^{-1}$). 
This DDT, in addition to ensuring an excellent fit to the observed SN rates, 
supports the hypothesis of a double-degenerate progenitor origin 
(i.e., a merger of two white dwarfs) for SNe Ia \citep{1984ApJ...277..355W}. 
The SNIa rate is shown with the red line in Fig.\ref{psi}.

The cosmic rates of the three classes of transients considered in this work are 
compared in Fig. \ref{psi}. 
$\psi_{\rm SGRB}(z)$ and $\psi_{\rm SNIa}(z)$ peak at a lower redshift than 
$\psi_{\star}(z)$ because of the delay between their formation as a binary and their merger. $\psi_{\rm LGRB}(z)$ peaks at a higher redshift because of the metallicity bias.

\subsection{Star formation factor} \label{sec:3.2}

In order to account for the preference of LGRBs and core-collapse SNe (Ib/Ic/IIp, CCSNe) to occur in regions characterized by a  high SFR, 
we define (similarly to \citealt{2015ApJ...810...41L}) the sSFR correction factor
\begin{equation}\label{eq:10}
f_{\rm sSFR} (R,z) =\frac{sSFR (R,z)}{sSFR (z)}
\end{equation}
that describes the fraction of the sSFR within the MW, $sSFR (R,z)$, with respect to the specific cosmic star formation 
rate of the Universe at the same epoch $sSFR (z)$. $sSFR(R,z)$ is estimated  
(\S \ref{sec:4.2}) through a  model \citep{2006MNRAS.366..899N,2009ApJ...690.1452N} that describes the 
evolution of the radial profile of the star formation and stellar surface density of the MW. 
The local cosmic specific star formation rate $sSFR(z)$ is defined as the ratio 
of the local star formation rate $\psi_{\star}(z)$ \citep{2014ARA&A..52..415M} and the 
average stellar density as a function of redshift 
$\rho_{\star}(z) = 10^{17.46-0.39z}$ M$_{\odot}$ Gpc$^{-3}$ \citep{2015ApJ...810...41L,2015MNRAS.447....2M}. As expected, when we combine eq. \ref{eq:8} and eq. \ref{eq:10}, the terms $\psi_{\star}(z)$ cancel out (i.e., the global cosmological evolution of the SFR is irrelevant when we consider a specific galaxy), only the present-day value $\psi_{\star}(0)$ counts as a normalization factor.

\subsection{Metallicity factor} \label{sec:3.3}

\medskip
In order to account for the preference of LGRBs to occur in low-metallicity environments, we define the correction factor 

\begin{equation}
f_{\rm Fe} (R,z) =\frac{\Theta_{\rm Z < Z_c} (R,z)}{\Theta_{\rm Z < Z_c}(z)}
\end{equation}
as the fraction of stars with a metallicity lower than $Z_c=0.4 Z_{\odot}$ (at any $R$ and $t$) in the MW divided by the fraction of stars with the same metallicity threshold in the local Universe. 
This definition is similar to what was adopted by \citet{2015ApJ...810...41L}, but we  consider for the first time the metallicity profile and its time evolution within the MW. 

As explained in $\sec{3.1.1}$,  $\Theta_{Z<Z_c}(z)$ is elided with the same term as in Eq. 10. Eq. \ref{eq:8} becomes 

\begin{equation}
    \mathcal{P} (d,z \mid R) = \frac{1}{M_{\star} (z)} \int_{S} \Sigma_{\star} (R,z) f_{\rm sSFR} (R,z) f_{\rm Fe} (R,z) \ da
.\end{equation}

The progenitor difference of SGRBs and LGRBs accounts for the preference of LGRBs to reside in environments of high star formation and low metallicity. Conversely, there is no evidence of a preference of SGRBs for environments of high star formation and low metallicity. For this reason, we assume 
$f_{\rm sSFR, SGRB}$ = 1 and $f_{\rm Fe,SSGRB}$=1. The same holds for SNIa. For CCSNe we assume $f_{\rm sSFR, CCSN}$ as calculated in Eq. \ref{eq:10} and $f_{\rm Fe,CCSNe}$=1 because for a progenitor with a mass $\le$ 40 M$_{\odot}$ , the formation of a SN is independent of metallicity \citep{2003ApJ...591..288H}. The rare case of SNe that originated in a progenitor with mass $\ge$ 40 M$_{\odot}$ does not change our results significantly.

In order to account for the errors on the parameters of the luminosity functions, we implemented for the distribution of GRBs durations, SNe energies, and rates a Monte Carlo simulation with 1000 realizations for each type of lethal event. We calculated the number of lethal events during a time interval of 500 Myr as the median value of the distribution of realizations. For each simulation we extracted $\rho$, $\alpha$, $\beta$, $E_{\rm SN}$,  $L_{b}$, and $\tau$. For the first four parameters we assumed that they follow Gaussian distributions with characteristic values as reported in Table \ref{tab:1} and  Table \ref{tab:2}. For  $L_{b}$ and $\tau$ we sampled  a log-normal distribution with characteristic values as reported in Table \ref{tab:1} and  \ref{tab:2}. 

\subsection{Collimated emission from GRBs}

In order to take the strong collimation of GRBs into account, we used the {\it \textup{observed rate}} and the {\it \textup{isotropic equivalent luminosity}} in our computations.  $n_0$ in eq. \ref{eq:6} is the observed GRB rate (i.e., not corrected for collimation) at the break isotropic equivalent luminosity $L_b$, hence it accounts for GRBs whose jets point toward the Earth. The true event rate, that is, the rate that includes the inferred large population of misaligned GRBs, can simply be calculated by dividing $n_0$ by the collimation factor $f_b=(1-\cos\theta_{\rm jet})$, where $\theta_{\rm jet}$ is the jet half-opening angle. In practice, $f_b$ gives the fraction of jets pointing at Earth. When we wish to calculate the rate of GRBs that can cause a lethal event, we would need to multiply the true rate by the collimation factor. Therefore the collimation factor cancels out. Explicitly, we have 

\begin{equation} \label{eq:3}
     \frac{d \mathcal{N}_{\rm MW}(R, z)}{d z}  = \int f_b \ \frac{\phi(L)}{f_b}  \ \psi_{\rm GRB}(z)  V_{\rm MW} (z) \mathcal{P} (d, z | R) \ dL. 
\end{equation}

\begin{table}
\centering
\begin{tabular}{ccc}
\hline
  \multicolumn{1}{c}{SN type}&
  \multicolumn{1}{c}{Rate (z=0)} &
  \multicolumn{1}{c}{Burst energy}  \\
  \ & \ $10^4 \ \rm Gpc^{-3} yr^{-1}$ &[erg] \\
\hline
 Ia & 2.2$\pm$ 0.3 [a] & 10$^{46 \pm 1}$ [c]\\
 Ibc & 2.6$\pm$ 0.4 [b] & 10$^{46 \pm 1}$ [d]\\
 IIp & 3.1 $\pm$ 0.5 [b] & 10$^{44 \pm 1}$ [e]\\
\hline
\end{tabular}
\medskip
\caption{Parameters for the populations of SNe: cosmic rate and released energy ($E_{\rm SN}$) for each SN type, as reported by \citet{2011AsBio..11..343M}. [a] \citet{2012PASA...29..447M}, [b] \citet{2011MNRAS.412.1441L}, [c] \citet{2009ApJ...705..483H}, [d] \citet{2008Natur.453..469S}, and [e] \citet{2008Sci...321..223S}}

\label{tab:2}
\end{table}

\subsection{Galaxy model} \label{sec:4.2}
In order to track the evolution and distribution of stellar surface density, SFR, and metallicity of the Galaxy, we used the inside-out formation model of \citet{2009ApJ...690.1452N}, which reproduces several observable properties of the present-day MW (Table \ref{tab:3}). In this model the authors incorporated the bulge, which instead was neglected in the 2006 version of the model \citep{2006MNRAS.366..899N}. In the following we summarize this galaxy model.

\citet{2009ApJ...690.1452N} assumed that the formation and evolution of MW proceeds in two phases. In the first phase (i.e., for cosmic time t<2.5 Gyr), the galaxy is still coupled to the hierarchical growth of the large-scale structure. In this phase, the infall gas loses its angular momentum more effectively as a result of shocks and tidal torques. The authors associated this phase of high infall rates with the formation epoch of the bulge: after $t_{\rm form}= 2.5$ Gyr, the baryonic mass (i.e., mainly gas during early stages) settles in the bulge with a steep exponential surface density profile,
\begin{equation}
    \Sigma_{\rm b} (R,z<z_{\rm form}) =  \Sigma_{0,b} (z_{\rm form}) \exp[-R/R_s(z_{\rm form})],
\end{equation}

with $\Sigma_{0,b} (z_{\rm form})=10^4$ M$_{\odot}$ pc$^{-2}$, while R$_b(z)=$0.6 kpc. At this epoch, the DM halo reaches its present-day virial velocity and evolves in isolation. Then, the long and quiescent assembly (i.e., at low infall rate) of the disk can start. \citet{2009ApJ...690.1452N} assumed that the baryonic mass in the disk, starting from this time, evolves with an exponential profile, 

\begin{equation}
    \Sigma_{\rm d} (R,z<z_{\rm form}) = \Sigma_{0,d} (z_{\rm form}) \exp[-R/R_d(z)],
\end{equation}

where the central surface density is fixed at $z_{\rm form}$ (400 M$_{\odot}$ pc$^{-2}$), while the scale lenght $r_d (z)$ evolves as a fraction $f_d$ of the virial radius of the halo, 
up to the present-day value of 3.6 kpc,

\begin{equation}
    R_d (z<z_{\rm form}) = f_d \frac{v_{vir} (z_{\rm form})}{10 H(z)}.
\end{equation}

In order to compute the surface stellar density $\Sigma_{\star} (R,z) $ and metallicity $Z(R,z)$ distributions, \citet{2006MNRAS.366..899N, 2009ApJ...690.1452N} adopted a simplified version of the chemical evolution recipe proposed by \citet{1975ApJ...201L..51O}, neglecting 
radial gas flows. The model considered the instantaneous metal injection from massive stars (K$_{\rm ins}$), the delayed injection form low-mass stars (K$_{\rm late}$) and the gas-infall rate ($\Sigma_{\rm IFR}$), and estimates the variation of surface density in gas and stars (see \citealt{2006MNRAS.366..899N} for details),
\begin{equation}
\begin{aligned}
    d\Sigma_{g} (R,z) = -\Sigma_{\rm SFR} (R,z) dz + K_{\rm ins} (R,z)dz + \\
    + K_{\rm late}(R,z) dz + \Sigma_{\rm IFR}(R,z) dz,
\end{aligned}
\end{equation}

\begin{equation}
\begin{aligned}
    d\Sigma_{\star} (R,z) = +\Sigma_{\rm SFR} (R,z) dz - K_{\rm ins} (R,z)dz + \\
    - K_{\rm late}(R,z) dz. 
\end{aligned}
\end{equation}

\begin{figure}
   \centering
   \includegraphics[scale=0.45]{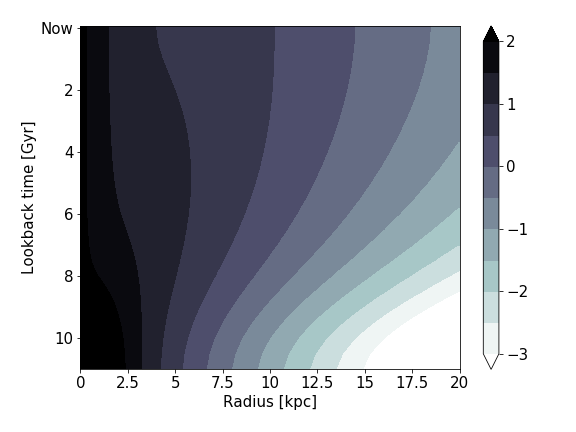}
   \caption{Star formation surface density as a function of the galactocentric radius $R$ and lookback time. The color-coding (left color bar) is on a logarithmic scale and in units of M$_{\odot} {\rm pc}^{-2} \rm Gyr^{-1} $.}
    \label{portion1}%
    \end{figure}

In order to account for the different star formation histories of bulge and disk, a formulation derived by  \citet{1998ApJ...498..541K} based on the local dynamical time (rotation period) of the system is assumed, that is,  
\begin{equation}
\Sigma_{\rm SFR} (R,z) = \epsilon \frac{\Sigma_{\rm g} (R,z) }{\tau (R,z)},
\end{equation}
where  $\epsilon=0.1$ is the star formation efficiency and $\tau(R,z)=2 \pi R/v(R,z)$ the orbital period, with $v$ as the local circular velocity. Because of smaller radii, higher gas surface density, and higher circular velocity, the star formation in the bulge is initially very high (Fig.\ref{portion1}), and most of its initial gas reservoir is then rapidly consumed.  

\begin{figure}
   \centering
   \includegraphics[scale=0.45]{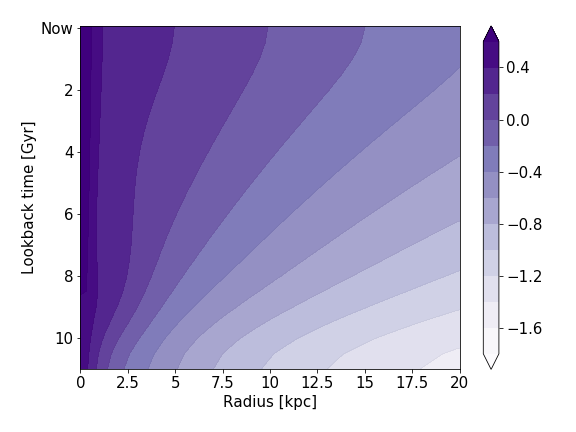}
   \caption{Metallicity as a function of the galactocentric radius $R$ and lookback time. The shaded regions (left color code) represent $Z/Z_{\odot}$ on a logarithmic scale.} 
    \label{portion2}%
    \end{figure}

Fig.\ref{portion1} shows the SFR surface density as a function of the position within the Galaxy (i.e., galactocentric radius $R$) versus lookback-time. The density contours clearly show the increase in SFR surface density from the inner part of the Galaxy toward the peripheral regions (inside-out star formation): while the innermost part of the Galaxy shows little evolution of the star formation after the early stage, the outskirts experienced an increase by several orders of magnitude. Fig. \ref{portion2} shows the evolution over cosmic time of the radial profile of the metallicity (on a lorgarithmic scale; left color bar). Consistently with the increase in star formation in the MW outskirts, the metallicity at larger distances from the Galaxy center also increased over the last Gyr. 

\begin{table}
\centering
\begin{tabular}{cc}
Propriety & Value \\
\hline
M$_{\star}$ & $5 \times 10^{10}$ M$_{\odot}$ \\
M$_{\rm gas}$ & $1 \times 10^{10}$ M$_{\odot}$ \\
$\Sigma_{\star, \odot}$ & $35$ \ M$_{\odot} \rm pc^{-2}$ \\
$\Sigma_{\rm g, \odot}$ & $15$ \ M$_{\odot} \rm pc^{-2}$ \\
SFR & $3$ \ M$_{\odot} \rm yr^{-1}$ \\
\hline
\end{tabular}
\medskip
\caption{Present-day Milky Way proprieties reproduced by model: total mass in stars, total mass in gas, stellar surface density at the solar radius, gas surface density at the solar radius, and global SFR.}
\label{tab:3}
\end{table}

\subsection{Planetary formation} \label{sec:4.3} 
TPs are typically defined as having a solid surface with radius and mass in the ranges $\rm 0.5-2.0R_{\Earth}$ and $\rm 0.5-10 M_{\Earth}$, respectively (where $R_{\Earth}$ and $M_{\Earth}$ are the radius and mass, respectively, of Earth). TPs might develop  habitable conditions \citep{2014A&A...561A..41A}. In order to  estimate the surface number density of TPs within the MW as a function of cosmic time and galactocentric distance, we adopted the model of \citet{2016ApJ...833..214Z}. As simulations and observations (radial velocity and transit surveys) suggest, close-orbit giants form in metal-enriched environments, while in a very low metallicity environment, planet formation is inhibited. The model, based on the assumption that close-orbit giants destroy the prospect of harboring TPs, assumes that the probabilities of forming TPs ($P_{\rm FTP}$) and close-orbit giants ($P_{\rm FG}$) are functions of the metallicity of the environment. Following \citet{2004Sci...303...59L}, the probability for a star to harbor a TP is defined as

\begin{equation}
    P_{\rm HTP} = P_{\rm FTP} (1-P_{\rm FG})
.\end{equation}

\citet{2016ApJ...833..214Z} approximated the probability of forming close-orbit giants as a function of the metallicity $[Fe/H]$ and of the stellar mass $M_{\star}$ \citep{2014ApJ...791...54G}, 

\begin{equation}
P_{\rm FG}([Fe/H], M_{\star})=f_0 10^{a [Fe/H]} M_{\star}^b
,\end{equation}
where $f_0=0.07$ is a constant factor  and the parameter values are $a=1.8(1.06)$ for FGK (M) stars and $b=1$ \citep{2014ApJ...791...54G}. The probability of forming TPs is \citep{2016ApJ...833..214Z}
\begin{equation}
    P_{\rm FTP} = f_{\rm TP} k(Z)
,\end{equation}
with $f_{\rm TP}=0.4 (1)$ for FGK (M) stars. $k(Z)$ is a function with a cutoff at low-metallicity values,
\begin{equation*}
k(Z) = \begin{cases}
0 &\text{if [Fe/H] $<$ -2.2}\\
\frac{Z-0001}{0.001-0.0001} &\text{if -2.2 $\le$ [Fe/H] $\le$ -1.2 }\\
1 &\text{if [Fe/H] $>$ -1.2.}
\end{cases}
\end{equation*}

Combining these equations with the Galaxy model of \citet{2009ApJ...690.1452N} (\S 3.2), we 
can compute $P_ {\rm HTP} (R,t)$ accounting for the metallicity radial distribution and its evolution within the Galaxy. The number surface density of TPs as function of time in the MW was computed using the  star formation surface density derived in \S 3.2 and assuming a Salpeter initial mass function: we computed the fraction of M stars $f_{\rm M}$ (with masses in the range 0.1-0.6 $M_{\Sun}$) and FGK stars $f_{\rm FGK}$ (with masses in the range 0.6-1.2 $M_{\Sun}$). Assuming an average mass for M stars of 0.35 $M_{\Sun}$ and 0.9 $M_{\Sun}$ for FGK stars, we derived the number surface density of TPs around M  and FGK stars with the following equation:
\begin{equation}
    \Sigma_{\rm TP}(R,z) = \int_{z_{\rm form}}^{z} \frac{f \ \Sigma_{\rm SFR} (R,z) P_{\rm HTP} (R,t)}{<M>} dz
,\end{equation}

where $z_{\rm form} = 3$ is the formation redshift of the MW in the model of \citet{2009ApJ...690.1452N}.

\section{Results}\label{sec:5} 
\begin{figure}
   \centering
   \includegraphics[scale=0.45]{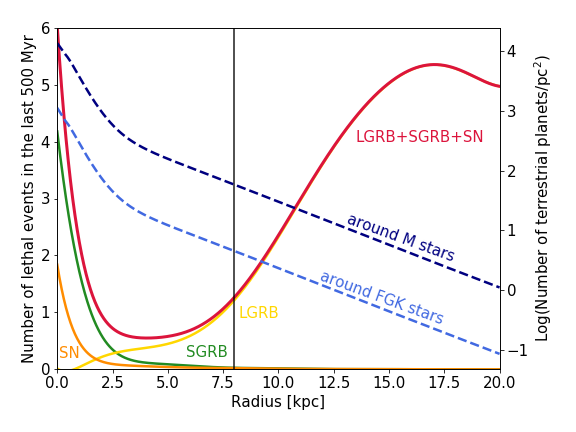}
   \caption{Number of lethal astrophysical transient events (solid red line) in the past 500 Myr as a function of distance from the Galactic center. The individual contributions of SGRBs (solid green line), LGRBs (solid yellow line) and SNe (solid orange line) are shown. The surface number density of TPs  around M stars (dashed dark blue line) and around FGK stars (dashed light blue line) are reported (left vertical axis). The vertical solid black line at 8 kpc marks the position of the Solar System at which the total number of lethal events (predominantly LGRBs) is $\sim$1.3.}
    \label{500}%
    \end{figure}

Figure \ref{500} shows the number of lethal astrophysical transient events in the past 500 Myr as a function of distance from the Galactic center. At the position of Earth (8 kpc, vertical solid black line), the number of lethal events (predominantly LGRBs) is about one to two within the past 500 Myr. This agrees with the hypothesis ascribing the Ordovician mass-extinction event to a LGRB \citep{2005GeoRL..3214808M}. The minimum in the solid red line (i.e., LGRBs+SGRBs+SNe) identifies a region between $\sim$2 and $\sim$8 kpc from the Galaxy center in which life in the past 500 Myr might have experienced $\le$1 lethal events. In the outskirts (R$>$10 kpc) and in the center (R$<$1.5 kpc) of the MW the number of lethal event is $>$2. In the outskirts of the Galaxy, the predominant lethal events  are LGRBs because of the low metallicity in the environment, while in the star-forming of the center, SGRBs and SNe occur predominantly. The outskirts of the MW are not favored to host life because only a few TPs are located there (dashed lines) and because lethal LGRBs are very frequent. 

\begin{figure*} 
   \centering
   \includegraphics[scale=0.55]{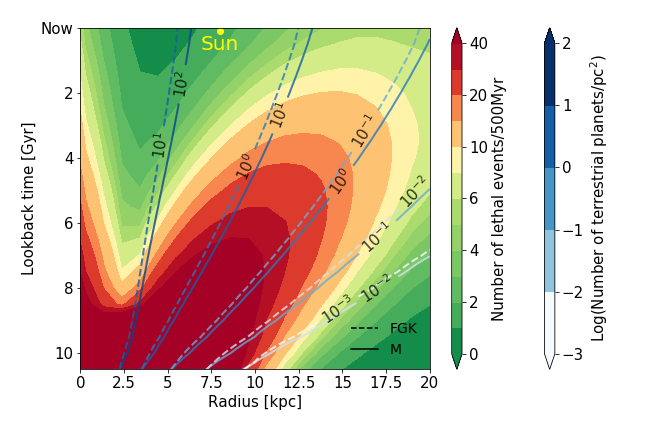}
   \caption{Number of lethal events per bins of 500 Myr as a function of the Galactic radius and loockback time. Shaded contours (corresponding to the green, yellow, and red color bar) show the number  of  lethal events (SSGRBs+LGRBs+SNs) per 500 Myr as a function of the distance from the Galactic center (x-axis) throughout the cosmic history of the Milky Way (y-axis). The line contours (corresponding to the blue-scale color bar) show the surface number density of TPs around M stars (solid lines) and  FGK stars (dashed lines). The current position of the Solar System is marked by the yellow dot. The Galaxy disk portion that extends $\sim$2 and $\sim$8 kpc away from the Galaxy center, where the TP density is relatively higher, represents the place in which life in the past 4 Gyr might have experienced less frequent major damages as a result of ozone depletion induced by transient astrophysical events. } 
    \label{total}%
    \end{figure*}
   
Figure \ref{total} shows the number of lethal events (SGRBs+LGRBs+SNe) in bins of 500 Myr as a function of the distance from the Galactic center (x-axis) throughout the cosmic history of the Milky Way (y-axis). The line contours (corresponding to the blue-scale color bar) show the surface number density of TPs around M stars (solid lines) and FGK stars (dashed lines). 

\begin{figure}
   \centering
   \includegraphics[scale=0.5]{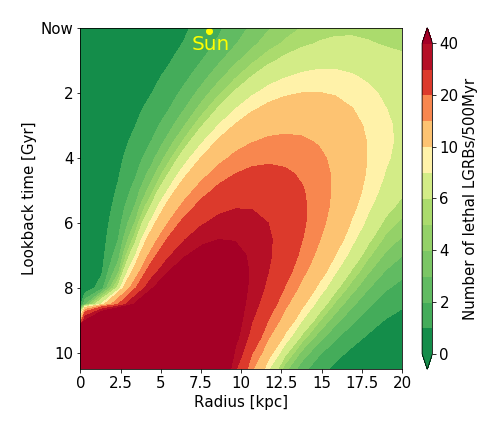}
   \caption{Number of lethal LGRBs per bins of 500 Myr as a function of the Galactic radius and lookback time.} 
    \label{long}%
\end{figure}

\begin{figure}
   \centering
   \includegraphics[scale=0.5]{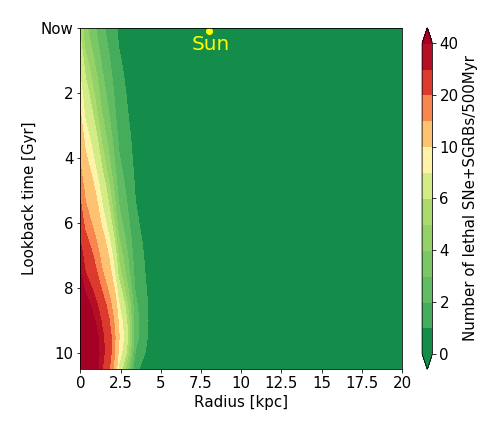}
   \caption{Number of lethal SGRBs and SNe per bins of 500 Myr as a function of the Galactic radius and lookback time.} 
    \label{short}%
\end{figure}

The individual contributions of SNe, SGRBs and LGRBs to Fig.\ref{total} are show in Fig.\ref{long} and Fig.\ref{short}. SGRBs and SNe (Fig.\ref{short}) are concentrated in the central regions of the Galaxy because the stellar density is high and their occurrence is independent of the metallicity of the environment. On the other hand, the incidence of LGRBs as lethal events (Fig.\ref{long}) develops throuhout the MW history with  an inside-out pattern: They dominate the rate of lethal events at the early stages of the MW evolution in the central regions, where most of the stars are formed, but are progressively suppressed by the increase in metallicity, and they become more prominent in the outskirts where the star formation is relatively higher and the metal pollution is not yet high. 

There are two  so-called green valleys within the MW disk and its cosmic history. The first is in the  outer regions of the disk (i.e., $R>12$ kpc) that experienced a relatively low incidence of lethal events during the first 6 Gyr of Galaxy evolution. This region has the lowest TP surface density, however. Starting about 6 Gyr ago, LGRBs became the dominant lethal sources for life within the MW because of their energetics, with an increasing number of lethal events toward the Galaxy periphery (red to orange contours in Fig. \ref{long}). This is due to the increased conversion rate of gas relatively little polluted by metal into massive stars in the outer regions of the MW. This global trend determined the formation of an increasingly larger, safer region of the MW located at intermediate galactocentric distances $R\in(2-8)$ kpc. In this second green valley, the main contribution of lethal events is still due to LGRBs, but the increase in metallicity due to the intense star formation suppresses the incidence of LGRBs. The higher density of TPs in this region makes it the most favorable region in the Galaxy for the development and resilience of life to ozone depletion induced by transient astrophysical events in the last 4 Gyr. In general, the early stages of the MW evolution (from its formation until 6 billion years ago) witnessed extremely poor conditions for life development because lethal events occurred throughout almost the entire Galactic disk ($R<10-12$ kpc), in which TPs are present in a considerable number.

\begin{figure}
   \centering
   \includegraphics[scale=0.5]{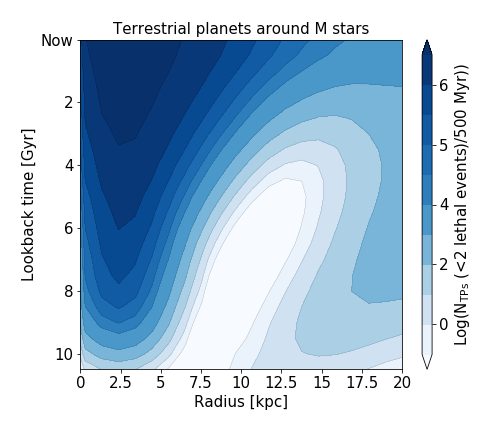}
   \caption{Number of TPs around M stars that experienced fewer than two lethal events in 500 Myr. The number is obtained from the surface density of planets integrated in annuli of constant width ($=1$ pc).}
    \label{MTPS}%
\end{figure}
    
\begin{figure}
   \centering
   \includegraphics[scale=0.5]{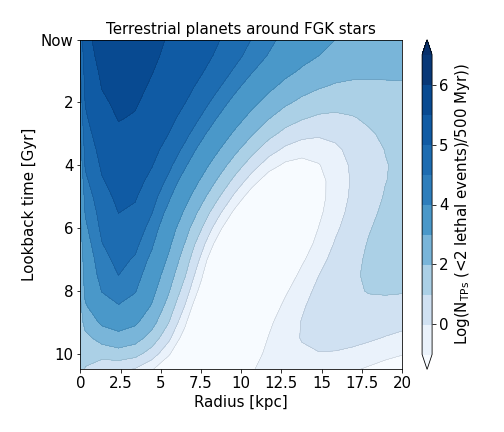}
   \caption{Same as fig.~\ref{MTPS}, but for FGK stars.}
    \label{FGKTPS}%
\end{figure}

Although it is not clear how many mass extinctions can prevent the habitability of a planet, based on what happened to Earth, we consider that one lethal event within 500 Myr at most is a conservative tolerable rate of lethal events for life as we know it to survive. The number of terrestrial potentially habitable planets orbiting M and FGK stars as a function of Galactocentric distance and cosmic time is shown in fig.~\ref{MTPS} and fig.~\ref{FGKTPS}, respectively. We first calculated the probability of having fewer than two life-threatening events in 500 Myr as a Poisson distribution, with mean the number of lethal events/500 Myr shown in Fig \ref{total}, and then multiplied by the number of TPs around M and FGK stars, obtained by integrating the planet surface density in annuli of constant width $=1$ pc.

\section{Discussion} \label{sec:6}
We measured the lethal power of the three classes of transients considered from their energetics, event rates, and the surface density of TPs in the MW. We found that LGRBs are the dominant life-threatening transients for planets at $R>2$ kpc throughout the entire cosmic history of the  MW (Fig.\ref{long}). This  is  mainly  due  to  their  high  energy  ($10^{51-54}$ ergs, isotropic equivalent), which compared to the energy of SNe compensates for their lower intrinsic rate. Instead, SGRBs (which compete with LGRBs in terms of energetics) and SNe dominate the counts of hazardous transients in the central regions of the MW, especially during the earliest evolution (Fig. \ref{short}), because their occurrence is mainly driven by stellar density and SFR. We discuss the effect of the many assumptions we made in our analysis in this section. 

\subsection{Transients}
The fluence threshold we considered harmful for life (i.e., $\gtrsim 10^{8}$ erg cm$^{-2}\,\,\equiv 100$ kJ m$^{-2}$) would induce a $\simeq 90$\% depletion of the ozone layer. This reduction can trigger mass extinctions \citep{2005ApJ...622L.153T}.
The choice of the threshold value affect the estimate of the rate of lethal events, but not on their spatial and temporal trends within the MW. If we were to consider the lower threshold fluence of 10 kJ m$^{-2}$ (able to induce a 68\% depletion of the ozone layer), for example, then there would be lethal events at larger distances, hence increasing the total rate at any $R$ and at any time. In contrast, a higher threshold fluence of 1000 kJ m$^{-2}$ (98\% depletion of the ozone layer), for instance, would select only the most powerful transients that explode closer to any given planet. These events are relatively rare given the corresponding steep luminosity function of LGRBs and SGRBs.
The threshold adopted in our study is higher than the lethal dose typically considered for Earth biota, that is, 10$^{5}$ erg cm$^{-2}$ for eucaryotic multicellular lifeforms \citep{2002ApJ...566..723S} and 10$^{7}$ erg cm$^{-2}$ for prokaryotic microbes  \citep{2017NatSR...716626B}. This allowed us to extend our results also to planets without an ozone layer. 

The lethal impact of SGRB, which dominate the central regions of the Galaxy (Fig.\ref{short}), was computed assuming the luminosity function derived by \citet{2016A&A...594A..84G}. Alternatively, if we had considered a steeper faint-end of the SGRB luminosity function  (such as in \citealt{2015MNRAS.448.3026W}), the rate of low-luminosity SGRBs would be higher. These events, however, are characterized by relatively short hazard distances (Eq. \ref{eq:4}) and our results would be practically unaltered. 

We linked the LGRB and the formation rate of TPs to the metallicity evolution within the MW. We considered that LGRBs preferably form in evironments with $Z<0.4Z_{\odot}$ \citep{2017MNRAS.469.4921B, 2018NewA...65...73B}. Assuming a different metallicity threshold (Tab. \ref{tab:4}) mainly affects the rate of lethal LGRBs in the outer regions of the Galaxy, where the metallicity is indeed close to the  threshold we did adopt.  

\begin{table}
\centering
\begin{tabular}{cccc}
\hline
  \multicolumn{1}{c}{} &
  \multicolumn{1}{c}{0.3 Z$_{\odot}$} &
  \multicolumn{1}{c}{0.4 Z$_{\odot}$} &
  \multicolumn{1}{c}{0.5 Z$_{\odot}$}  \\
\hline
 8 kpc & 1.5 & 1.2 & 1.0 \\
 17 kpc & 2.9 & 5.3 & 7.4 \\
\hline
\end{tabular}
\medskip
\caption{Number of lethal LGRBs at 8 kpc and at 17 kpc in the past 500 Myr considering different metallicity thresholds.} 

\label{tab:4}
\end{table}

We did not take the effect of cosmic rays into account. In a recent work, \citet{2020SciA....6A.768M} suggested that the moderate extinction event at the end of the Devonian period (Hangenberg event, $\sim$360 Myr) was associated with ozone depletion and a consequent higher solar UV-B radiation on Earth. This may explain the discovery of malformed land-plant spores in terrestrial Devonian-carboniferous boundary sections from East Greenland. Although a GRB cannot be excluded as the cause of this extinction event, \citet{2020arXiv200701887F} proposed cosmic rays, accelerated in a nearby SN and magnetically confined inside the SN remnant for $\sim$100 kyr, as the trigger of the ozone-layer depletion followed by the Devonian extinction. 

\subsection{UV flash}
An effect we did not explicitly consider is the effect caused by UV flashes. {\citet{2007IJAsB...6...19G} and \citet{2009Ap&SS.321..161M}} suggested that the UV flash resulting from the transmission of the $\gamma$-rays  through the atmosphere could have important short-term effects on life forms, especially for planets with
thin atmospheres. In contrast,  
\citet{2009IJAsB...8..183T} argued that only long-term and extended effects on Earth-like life forms caused by ozone-layer depletion can trigger biological catastrophes of the type we considered here. A UV flash is indeed a very short-lived phenomenon, occurring on the timescale of the $\gamma$--ray event ($\sim$1 s and 10 s for SGRB and LGRB, respectively), and it is localized on half of the planet surface. However, a short but strong kick could be much more harmful than constant pressure over years,  such as that caused by the depletion of the ozone layer.  Moreover, a UV flash (1-10 s) and ozone depletion (10 yr) are essentially instantaneous compared to the ecological timescales of recovery after extinctions \citep{2000Natur.404..177K}.

\citet{2020arXiv201108433R} showed that a UV flash can reduce a significant fraction (20-60\%) of phytoplankton biomass, the very basis of the ocean food chain and oxygen production. To estimate the UV flash effect, they considered an event able to hit the top of the atmosphere with a $\gamma$-ray fluence of 100 kJ/m$^2$. Because this is the same critical fluence we assumed in our work,
this effect of a UV flash can be added to that of the ozone-layer depletion. If a lower $\gamma$-ray fluence at the top of the atmosphere produces an UV flash that can trigger a mass extinction, the total rate of lethal events increases, but their spatial and temporal distribution within the MW (Fig. \ref{total}) remains unchanged.

\subsection{Planets}
As suggested by \citet{2020Icar..35214025O}, planets forming early on in the MW history tend to have low Fe/Si ratios and thus strong activity of the tectonic plates, which appears to be an important factor for the development of life as we know it. However, as we showed, planets were subject to a very high rate of possibly lethal events at the early stages of
the Galaxy.

The adopted planet formation model \citep{2016ApJ...833..214Z} implies that gas giants with orbital period $\lesssim 2$ yr \citep[i.e., warm and hot Jupiters, see][]{2014ApJ...791...54G} hinder the presence of TPs. However, this assumption is likely to overstate the effect of close-orbit giants on TPs, as there are examples of warm Jupiters in systems with 
rocky members \citep{2012PNAS..109.7982S, 2016ApJ...825...98H}. Our estimate of the surface density of TPs should therefore be regarded as a lower limit.

\subsection{The Galaxy}
The Naab \& Ostriker galaxy model does not take any azimuthal inhomogeneities into account. The innermost regions could be even more unfavorable to life if a galactic bar were present that would enhance the  rate of lethal events (Fig \ref{total}, R< 4 kpc). The same applies to the regions contained within the spiral arms. 

Neglecting the distinction between thin- and thick-disk stars is highly justified in our study, as the thick disk accounts for $\simeq $10\% of the stars in our galaxy, predominantly low-mass stars that do not produce the dominant lethal events (CCSNe and LGRBs). Moreover, the range (a few kiloparsec) of the dominant lethal events in the disk (led by LGRBs) is larger than the average thickness of the thick disk, and the segregation in metallicity between thick- and thin-disk stars (Yan et al. 2019) is likely marginal in the probability of planet formation.

We did not consider other aspects that were investigated in recent works in the context of Galactic habitability. For example, the combined action of tidal disruption events \citep{2020MNRAS.tmp.2455P} and the active phase of the central black hole  \citep{2017NatSR...716626B} further reduce the habitability of the bulge. Moreover, a detailed evaluation of the habitability of the bulge should also account for the relatively higher probability of stellar encounters that can perturb planetary orbital configuration, but on the other hand,  also  favor lithopanspernia \citep{1988Natur.332..687M,2010SSRv..156..239W}. 

\medskip

\section{Conclusions}
\label{sec:conclusion}
We have investigated the impact of the most energetic transient events on planetary habitability inside the MW and throughout its cosmic history. We considered LGRBs, SGRBs, and SNe (Ia, Ibc, and IIp). These are the most energetic transients and can irradiate a planetary atmosphere with a $\gamma$-ray fluence $\gtrsim 100$ kJ m$^{-2}$ (10$^{8}$ erg cm$^{-2}$). 
Our main results are summarized below. 

We confirmed that at the position of Earth, one LGRB may have occurred within the past 500 Myr. This may have played a leading role in the Ordovician mass extinction. 

We demonstrated that the evolutionary pressure due to the considered astrophysical events is not negligible during the evolution of our Galaxy. The safest zone in the past 500 Myr is within about 2-8 kpc. Different from what has been claimed in recent works, we find that the outskirts are not favored to host life because they host few TPs (dashed lines in Fig.\ref{500}) and many lethal LGRBs, which hinders the emergence of a long-lasting biosphere. SGRBs and SNe are the dominant lethal events only in the central regions of the Galaxy. Searches for exoplanets harboring life forms probably will have more chances of success in the direction of the Galactic center, within 5 kpc from the Sun, because of the combined effect of a high density of TPs (dashed line in Fig.\ref{500}) and of the low occurrence of lethal transients (solid red line in Fig.\ref{500}). 

We identify two  green valleys within the MW disk and its cosmic history (Fig.\ref{total}). The first is located in the outskirts of the galactic disk (i.e., $R>12$ kpc). These regions experienced a relatively low incidence of lethal events during the first 6 Gyr of the Galactic evolution. However, the probability that life would emerge is hampered by the the low TP surface density. The other green valley is located at intermediate galactocentric distances $R\in(2-8)$ kpc. Here the dominant lethal transients were  LGRBs until 7-6 Gyr, but later on, the progressive increase in metallicity due to the intense star formation suppressed  the lethal incidence of LGRBs. The higher density of TPs in this region makes it the most favorable place of the Galaxy for the resilience of life to mass extinction induced by transient astrophysical events over the last $\sim$4 Gyr. The inner region of the MW hosts most of the planets with the lowest probability to be at danger distance from more than one lethal event every 500 Myr (Fig \ref{MTPS} and Fig \ref{FGKTPS}). This is due to the high stellar density and high metallicity, which inhibits the formation of LGRBs. This result is in agreement with \citet{2018arXiv180207036G}.

In conclusion,  the most powerful cosmic explosion jeopardized life within most of the Milky Way in the past, but no longer.  In particular, until $\sim$6 Gyr ago, the entire Galaxy was very frequently sterilized by transient events. At the early stage of the Galaxy evolution, life as we know it must have been more resilient to high radiation irradiance in order to survive.  When we exclude the green valley at the bottom right side of Fig. \ref{total}, which has the lowest TP surface density, the Galaxy overall is a safer place to live on a planet today than it was in the past. When we assume that the Sun did not significantly migrate along the galactocentric radius during its lifetime, the Earth, from its birth until today,  experienced an ever lower rate of potential mass-extinction events and gradually became an increasingly safer place.

Finally, we note that the very existence of life on planet Earth today demonstrates that mass extinctions do not necessarily preclude the possibility of complex life development. On the contrary, mass extinctions occurring at the right pace could have played a pivotal role in the evolution of complex life forms on our home planet \citep{1985IAUS..112..223S,raup,jablonski,2012Geo....40..731K,stroud}.

\begin{acknowledgements} 
R.S. acknowledges the Brera Observatory for the kind hospitality during the completion of this work. The authors thank F. Borsa for useful discussions. Funding support is acknowledged from Prin-INAF 1.05.01.88.06 “Towards the SKA and CTA era: discovery, localisation, and physics of transient sources”; 1.05.06.13 Premiale 2015 “FIGARO”; Accordo Attuativo ASI-INAF n.2017-14-H.0. We acknowledge support from PRIN-MIUR 2017 (grant 20179ZF5KS).
\end{acknowledgements}


\begin{thebibliography}{}
\bibitem[Abbott et al.(2017a)]{2017PhRvL.119p1101A} Abbott, B.~P., Abbott, R., Abbott, T.~D., et al.\ 2017, \prl, 119, 161101
\bibitem[Abbott et al.(2017b)]{2017PhRvL.119n1101A} Abbott, B.~P., Abbott, R., Abbott, T.~D., et al.\ 2017, \prl, 119, 141101
\bibitem[Alibert(2014)]{2014A&A...561A..41A} Alibert, Y.\ 2014, \aap, 561, A41

\bibitem[Balbi \& Tombesi(2017)]{2017NatSR...716626B} Balbi, A. \& Tombesi, F.\ 2017, Scientific Reports, 7, 16626
\bibitem[Barris et al.(2004)]{2004ApJ...602..571B} Barris, B.~J., Tonry, J.~L., Blondin, S., et al.\ 2004, \apj, 602, 571
\bibitem[Bertelli et al.(1994)]{1994A&AS..106..275B} Bertelli, G., Bressan, A., Chiosi, C., et al.\ 1994, \aaps, 106, 275
\bibitem[Bignone et al.(2017)]{2017MNRAS.469.4921B} Bignone, L.~A., Tissera, P.~B., \& Pellizza, L.~J.\ 2017, \mnras, 469, 4921
\bibitem[Bignone et al.(2018)]{2018NewA...65...73B} Bignone, L.~A., Pellizza, L.~J., \& Tissera, P.~B.\ 2018, \na, 65, 73
\bibitem[Botticella et al.(2008)]{2008A&A...479...49B} Botticella, M.~T., Riello, M., Cappellaro, E., et al.\ 2008, \aap, 479, 49
\bibitem[Campana et al.(2006)]{2006Natur.442.1008C} Campana, S., Mangano, V., Blustin, A.~J., et al.\ 2006, \nat, 442, 1008
\bibitem[Cappellaro et al.(1999)]{1999A&A...351..459C} Cappellaro, E., Evans, R., \& Turatto, M.\ 1999, \aap, 351, 459
\bibitem[Dar et al.(1998)]{1998PhRvL..80.5813D} Dar, A., Laor, A., \& Shaviv, N.~J.\ 1998, \prl, 80, 5813
\bibitem[Daigne et al.(2006)]{2006MNRAS.372.1034D} Daigne, F., Rossi, E.~M., \& Mochkovitch, R.\ 2006, \mnras, 372, 1034
\bibitem[D'Avanzo et al.(2014)]{2014MNRAS.442.2342D} D'Avanzo, P., Salvaterra, R., Bernardini, M.~G., et al.\ 2014, \mnras, 442, 2342
\bibitem[Fields et al.(2020)]{2020arXiv200701887F} Fields, B.~D., Melott, A.~L., Ellis, J., et al.\ 2020, arXiv:2007.01887
\bibitem[Firmani et al.(2004)]{2004ApJ...611.1033F} Firmani, C., Avila-Reese, V., Ghisellini, G., et al.\ 2004, \apj, 611, 1033
\bibitem[Fruchter et al.(2006)]{2006Natur.441..463F} Fruchter, A.~S., Levan, A.~J., Strolger, L., et al.\ 2006, \nat, 441, 463

\bibitem[Gaidos \& Mann(2014)]{2014ApJ...791...54G} Gaidos, E. \& Mann, A.~W.\ 2014, \apj, 791, 54
\bibitem[Galama et al.(1998)]{1998Natur.395..670G} Galama, T.~J., Vreeswijk, P.~M., van Paradijs, J., et al.\ 1998, \nat, 395, 670
\bibitem[Galante \& Horvath(2007)]{2007IJAsB...6...19G} Galante, D. \& Horvath, J.~E.\ 2007, International Journal of Astrobiology, 6, 19. doi:10.1017/S1473550406003545
\bibitem[Gallazzi et al.(2008)]{2008MNRAS.383.1439G} Gallazzi, A., Brinchmann, J., Charlot, S., et al.\ 2008, \mnras, 383, 1439
\bibitem[Gehrels et al.(2003)]{2003ApJ...585.1169G} Gehrels, N., Laird, C.~M., Jackman, C.~H., et al.\ 2003, \apj, 585, 1169

\bibitem[Ghirlanda et al.(2016)]{2016A&A...594A..84G} Ghirlanda, G., Salafia, O.~S., Pescalli, A., et al.\ 2016, \aap, 594, A84
\bibitem[Gowanlock et al.(2011)]{2011AsBio..11..855G} Gowanlock, M.~G., Patton, D.~R., \& McConnell, S.~M.\ 2011, Astrobiology, 11, 855
\bibitem[Gowanlock(2016)]{2016ApJ...832...38G} Gowanlock, M.~G.\ 2016, \apj, 832, 38
\bibitem[Gowanlock \& Morrison(2018)]{2018arXiv180207036G} Gowanlock, M.~G. \& Morrison, I.~S.\ 2018, arXiv:1802.07036
\bibitem[Guetta \& Piran(2005)]{2005A&A...435..421G} Guetta, D. \& Piran, T.\ 2005, \aap, 435, 421
\bibitem[Guetta \& Piran(2006)]{2006A&A...453..823G} Guetta, D. \& Piran, T.\ 2006, \aap, 453, 823
\bibitem[Guetta \& Della Valle(2007)]{2007ApJ...657L..73G} Guetta, D. \& Della Valle, M.\ 2007, \apjl, 657, L73
\bibitem[Hatano et al.(1997)]{1997MNRAS.290..360H} Hatano, K., Fisher, A., \& Branch, D.\ 1997, \mnras, 290, 360
\bibitem[Heger et al.(2003)]{2003ApJ...591..288H} Heger, A., Fryer, C.~L., Woosley, S.~E., et al.\ 2003, \apj, 591, 288
\bibitem[Herrmann \& Patzkowsky (2002)]{Herrmann2002} Herrmann, A.D., \& Patzkowsky, M.E. (2002) Astrobiology 2, 560-561
\bibitem[Herrmann et al.(2003)]{2003Geo....31..485H} Herrmann, A.~D., Patzkowsky, M.~E., \& Pollard, D.\ 2003, Geology, 31, 485

\bibitem[Hjorth et al.(2003)]{2003Natur.423..847H} Hjorth, J., Sollerman, J., M{\o}ller, P., et al.\ 2003, \nat, 423, 847

\bibitem[H{\"o}flich \& Schaefer(2009)]{2009ApJ...705..483H} H{\"o}flich, P. \& Schaefer, B.~E.\ 2009, \apj, 705, 483

\bibitem[Hopkins \& Beacom(2006)]{2006ApJ...651..142H} Hopkins, A.~M. \& Beacom, J.~F.\ 2006, \apj, 651, 142

\bibitem[Huang et al.(2016)]{2016ApJ...825...98H} Huang, C., Wu, Y., \& Triaud, A.~H.~M.~J.\ 2016, \apj, 825, 98. doi:10.3847/0004-637X/825/2/98

\bibitem[Jablonski (2001)]{jablonski} Jablonski, D. 2001, PNAS, 98, 5393-5398
\bibitem[Japelj et al.(2016)]{2016A&A...590A.129J} Japelj, J., Vergani, S.~D., Salvaterra, R., et al.\ 2016, \aap, 590, A129
\bibitem[Kasting et al.(1993)]{1993Icar..101..108K} Kasting, J.~F., Whitmire, D.~P., \& Reynolds, R.~T.\ 1993, \icarus, 101, 108
\bibitem[Kennicutt(1998)]{1998ApJ...498..541K} Kennicutt, R.~C.\ 1998, \apj, 498, 541
\bibitem[Kirchner \& Weil(2000)]{2000Natur.404..177K} Kirchner, J.~W. \& Weil, A.\ 2000, \nat, 404, 177. doi:10.1038/35004564
\bibitem[Kistler et al.(2009)]{2009ApJ...705L.104K} Kistler, M.~D., Y{\"u}ksel, H., Beacom, J.~F., et al.\ 2009, \apjl, 705, L104

\bibitem[Kopparapu et al.(2013)]{2013ApJ...765..131K} Kopparapu, R.~K., Ramirez, R., Kasting, J.~F., et al.\ 2013, \apj, 765, 131
\bibitem[Kouveliotou et al.(1993)]{1993ApJ...413L.101K} Kouveliotou, C., Meegan, C.~A., Fishman, G.~J., et al.\ 1993, \apjl, 413, L101
\bibitem[Krug \& Jablonski(2012)]{2012Geo....40..731K} Krug, A.~Z. \& Jablonski, D.\ 2012, Geology, 40, 731
\bibitem[Le Floc'h et al.(2006)]{2006ApJ...642..636L} Le Floc'h, E., Charmandaris, V., Forrest, W.~J., et al.\ 2006, \apj, 642, 636

\bibitem[Li \& Zhang(2015)]{2015ApJ...810...41L} Li, Y. \& Zhang, B.\ 2015, \apj, 810, 41
\bibitem[Li et al.(2011)]{2011MNRAS.412.1441L} Li, W., Leaman, J., Chornock, R., et al.\ 2011, \mnras, 412, 1441
\bibitem[Lineweaver et al.(2004)]{2004Sci...303...59L} Lineweaver, C.~H., Fenner, Y., \& Gibson, B.~K.\ 2004, Science, 303, 59
\bibitem[MacFadyen \& Woosley(1999)]{1999ApJ...524..262M} MacFadyen, A.~I. \& Woosley, S.~E.\ 1999, \apj, 524, 262

\bibitem[Madau \& Dickinson(2014)]{2014ARA&A..52..415M} Madau, P. \& Dickinson, M.\ 2014, \araa, 52, 415
\bibitem[Malesani et al.(2005)]{2005GCN..3087....1M} Malesani, D., Moretti, A., Romano, P., et al.\ 2005, GRB Coordinates Network, Circular Service, No. 3087, \#1 (2005), 3087
\bibitem[Maoz \& Mannucci(2012)]{2012PASA...29..447M} Maoz, D. \& Mannucci, F.\ 2012, \pasa, 29, 447

\bibitem[Marshall et al.(2020)]{2020SciA....6A.768M} Marshall, J.~E.~A., Lakin, J., Troth, I., et al.\ 2020, Science Advances, 6, eaba0768

\bibitem[Mart{\'\i}n et al.(2009)]{2009Ap&SS.321..161M} Mart{\'\i}n, O., Galante, D., C{\'a}rdenas, R., et al.\ 2009, \apss, 321, 161. doi:10.1007/s10509-009-0037-3

\bibitem[Melandri et al.(2012)]{2012A&A...547A..82M} Melandri, A., Pian, E., Ferrero, P., et al.\ 2012, \aap, 547, A82
\bibitem[Melosh(1988)]{1988Natur.332..687M} Melosh, H.~J.\ 1988, \nat, 332, 687
\bibitem[Melott et al.(2005)]{2005GeoRL..3214808M} Melott, A.~L., Thomas, B.~C., Hogan, D.~P., et al.\ 2005, \grl, 32, L14808
\bibitem[Melott \& Thomas(2011)]{2011AsBio..11..343M} Melott, A.~L. \& Thomas, B.~C.\ 2011, Astrobiology, 11, 343
\bibitem[Mortlock et al.(2015)]{2015MNRAS.447....2M} Mortlock, A., Conselice, C.~J., Hartley, W.~G., et al.\ 2015, \mnras, 447, 2

\bibitem[Naab \& Ostriker(2006)]{2006MNRAS.366..899N} Naab, T. \& Ostriker, J.~P.\ 2006, \mnras, 366, 899

\bibitem[Naab \& Ostriker(2009)]{2009ApJ...690.1452N} Naab, T. \& Ostriker, J.~P.\ 2009, \apj, 690, 1452
\bibitem[Nakar \& Gal-Yam(2005)]{2005AAS...20715803N} Nakar, E. \& Gal-Yam, A.\ 2005, American Astronomical Society Meeting Abstracts
\bibitem[Ostriker \& Tinsley(1975)]{1975ApJ...201L..51O} Ostriker, J.~P. \& Tinsley, B.~M.\ 1975, \apjl, 201, L51
\bibitem[O'Neill et al.(2020)]{2020Icar..35214025O} O'Neill, C., Lowman, J., \& Wasiliev, J.\ 2020, \icarus, 352, 114025

\bibitem[Pacetti et al.(2020)]{2020MNRAS.tmp.2455P} Pacetti, E., Balbi, A., Lingam, M., et al.\ 2020, \mnras, doi:10.1093/mnras/staa2535
\bibitem[Palmerio et al.(2019)]{2019A&A...623A..26P} Palmerio, J.~T., Vergani, S.~D., Salvaterra, R., et al.\ 2019, \aap, 623, A26
\bibitem[Pescalli et al.(2016)]{2016A&A...587A..40P} Pescalli, A., Ghirlanda, G., Salvaterra, R., et al.\ 2016, \aap, 587, A40

\bibitem[Pian et al.(2006)]{2006Natur.442.1011P} Pian, E., Mazzali, P.~A., Masetti, N., et al.\ 2006, \nat, 442, 1011
\bibitem[Piran \& Jimenez(2014)]{2014PhRvL.113w1102P} Piran, T. \& Jimenez, R.\ 2014, \prl, 113, 231102
\bibitem[Pescalli et al.(2016)]{2016A&A...587A..40P} Pescalli, A., Ghirlanda, G., Salvaterra, R., et al.\ 2016, \aap, 587, A40

\bibitem[Raup (1994)]{raup} Raup, D.M.\ 1994, PNAS, 91, 6758-6763
\bibitem[Ruderman(1974)]{1974Sci...184.1079R} Ruderman, M.~A.\ 1974, Science, 184, 1079

\bibitem[Richardson et al.(2002)]{2002AJ....123..745R} Richardson, D., Branch, D., Casebeer, D., et al.\ 2002, \aj, 123, 745

\bibitem[Rodr{\'\i}guez-L{\'o}pez et al.(2020)]{2020arXiv201108433R} Rodr{\'\i}guez-L{\'o}pez, L., Cardenas, R., Gonz{\'a}lez-Rodr{\'\i}guez, L., et al.\ 2020, arXiv:2011.08433
\bibitem[Salvaterra et al.(2012)]{2012ApJ...749...68S} Salvaterra, R., Campana, S., Vergani, S.~D., et al.\ 2012, \apj, 749, 68


\bibitem[Scalo \& Wheeler(2002)]{2002ApJ...566..723S} Scalo, J. \& Wheeler, J.~C.\ 2002, \apj, 566, 723
\bibitem[Schawinski et al.(2008)]{2008Sci...321..223S} Schawinski, K., Justham, S., Wolf, C., et al.\ 2008, Science, 321, 223

\bibitem[Sepkoski(1985)]{1985IAUS..112..223S} Sepkoski, J.~J.\ 1985, The Search for Extraterrestrial Life: Recent Developments, 112, 223

\bibitem[Soderberg et al.(2008)]{2008Natur.453..469S} Soderberg, A.~M., Berger, E., Page, K.~L., et al.\ 2008, \nat, 453, 469
\bibitem[Sparre et al.(2011)]{2011ApJ...735L..24S} Sparre, M., Sollerman, J., Fynbo, J.~P.~U., et al.\ 2011, \apjl, 735, L24

\bibitem[Spitoni et al.(2014)]{2014MNRAS.440.2588S} Spitoni, E., Matteucci, F., \& Sozzetti, A.\ 2014, \mnras, 440, 2588

\bibitem[Spitoni et al.(2017)]{2017A&A...605A..38S} Spitoni, E., Gioannini, L., \& Matteucci, F.\ 2017, \aap, 605, A38. doi:10.1051/0004-6361/201730545

\bibitem[Stanek(2003)]{2003ASAJ..113.1788S} Stanek, M.~J.\ 2003, Acoustical Society of America Journal, 113, 1788
\bibitem[Steffen et al.(2012)]{2012PNAS..109.7982S} Steffen, J.~H., Ragozzine, D., Fabrycky, D.~C., et al.\ 2012, Proceedings of the National Academy of Science, 109, 7982. doi:10.1073/pnas.1120970109

\bibitem[Stroud \& Losos (2016)]{stroud} Stroud J.T. \& Losos, J.N. 2016, Annual Review of Ecology, Evolution, and Systematics, 47, 507-532

\bibitem[Svensmark(2012)]{2012MNRAS.423.1234S} Svensmark, H.\ 2012, \mnras, 423, 1234

\bibitem[Thomas et al.(2005a)]{2005ApJ...622L.153T} Thomas, B.~C., Jackman, C.~H., Melott, A.~L., et al.\ 2005, \apjl, 622, L153

\bibitem[Thomas et al.(2005b)]{2005ApJ...634..509T} Thomas, B.~C., Melott, A.~L., Jackman, C.~H., et al.\ 2005, \apj, 634, 509

\bibitem[Thomas(2009)]{2009IJAsB...8..183T} Thomas, B.~C.\ 2009, International Journal of Astrobiology, 8, 183. doi:10.1017/S1473550409004509

\bibitem[Thorsett(1995)]{1995ApJ...444L..53T} Thorsett, S.~E.\ 1995, \apjl, 444, L53

\bibitem[Vergani(2018)]{2018MmSAI..89..175V} Vergani, S.~D.\ 2018, \memsai, 89, 175

\bibitem[Virgili et al.(2011)]{2011MNRAS.417.3025V} Virgili, F.~J., Zhang, B., Nagamine, K., et al.\ 2011, \mnras, 417, 3025
\bibitem[Vukoti{\'c} et al.(2016)]{2016MNRAS.459.3512V} Vukoti{\'c}, B., Steinhauser, D., Martinez-Aviles, G., et al.\ 2016, \mnras, 459, 3512

\bibitem[Wanderman \& Piran(2010)]{2010MNRAS.406.1944W} Wanderman, D. \& Piran, T.\ 2010, \mnras, 406, 1944
\bibitem[Wanderman \& Piran(2015)]{2015MNRAS.448.3026W} Wanderman, D. \& Piran, T.\ 2015, \mnras, 448, 3026
\bibitem[Webbink(1984)]{1984ApJ...277..355W} Webbink, R.~F.\ 1984, \apj, 277, 355
\bibitem[Wesson(2010)]{2010SSRv..156..239W} Wesson, P.~S.\ 2010, \ssr, 156, 239
\bibitem[Woosley et al.(1993)]{1993ApJ...411..823W} Woosley, S.~E., Langer, N., \& Weaver, T.~A.\ 1993, \apj, 411, 823
\bibitem[Woosley(1993)]{1993ApJ...405..273W} Woosley, S.~E.\ 1993, \apj, 405, 273
\bibitem[Woosley \& Heger(2006)]{2006ApJ...637..914W} Woosley, S.~E. \& Heger, A.\ 2006, \apj, 637, 914
\bibitem[Xu et al.(2013)]{2013ApJ...776...98X} Xu, D., de Ugarte Postigo, A., Leloudas, G., et al.\ 2013, \apj, 776, 98

\bibitem[Yasuda \& Fukugita(2010)]{2010AJ....139...39Y} Yasuda, N. \& Fukugita, M.\ 2010, \aj, 139, 39
\bibitem[Yoon et al.(2006)]{2006A&A...460..199Y} Yoon, S.-C., Langer, N., \& Norman, C.\ 2006, \aap, 460, 199
\bibitem[Zackrisson et al.(2016)]{2016ApJ...833..214Z} Zackrisson, E., Calissendorff, P., Gonz{\'a}lez, J., et al.\ 2016, \apj, 833, 214
\end{thebibliography}
\end{document}